\documentclass{article}
\usepackage{amsfonts}
\usepackage{amssymb}
\usepackage{epsfig}

\title{Three-body scattering in Poincar\'e invariant quantum mechanics
\thanks{
This work was performed in part under the
auspices of the U.~S.  Department of Energy under contract
No. DE-FG02-93ER40756 with Ohio University and
under contract No. DE-FG02-86ER40286 with the University of Iowa.
}
}
\author{W. N.  Polyzou \\
Department of Physics and Astronomy, \\
The University of Iowa, \\
Iowa City, Iowa 52242, USA 
\and
T. Lin, Ch. Elster \\
Institute of Nuclear and Particle Physics and \\
Department of Physics and Astronomy, \\
Ohio University, \\
Athens, OH 45701, USA 
\and
W. Gl\"ockle \\
Institut f\"ur Theoretische Physik II, \\
Ruhr-Universit\"at \\ 
Bochum, D-44780 Bochum, Germany
}

\begin{document}

\maketitle
\begin{abstract}

The relativistic three-nucleon problem is formulated by constructing a
dynamical unitary representation of the Poincar\'e group on the 
three-nucleon Hilbert space.  Two-body interactions are included that preserve 
the Poincar\'e symmetry,  lead to the same invariant two-body S-matrix 
as the corresponding non-relativistic problem,  and result in a
three-body S-matrix satisfying cluster properties.  The resulting 
Faddeev equations are solved by direct integration, without partial 
waves for both elastic and breakup reactions at laboratory energies up 
to 2 Gev.

\end{abstract}

\section{Introduction}

Energy scales near a GeV are interesting in nuclear physics because of
the relevance of sub-nuclear degrees of freedom.  Some experimental
manifestations of these degrees of freedom are the appearance of
baryon resonances and the opening of meson production channels.
Poincar\'e invariance is required for a consistent treatment of the
dynamics at these energies.  For a limited energy range the number of
relevant degrees of freedom should be small enough for a few-body
treatment.  A first step in modeling few-body systems at these
energies is to demonstrate the feasibility of solving Faddeev
equations in a Poincar\'e invariant quantum theory and understanding the 
differences in the predictions of the relativistic and 
non-relativistic theories.

In Poincar\'e invariant quantum mechanics the dynamics is given by a
dynamical representation of the Poincar\'e group \cite{1}.  The
construction of this representation is done in four steps.  The first
step is to choose a basis for single particle irreducible
representations of the Poincar\'e group.  In this work the irreducible
representation space is taken as the space of square integrable
functions of the linear momentum and z-component of the canonical
spin.  The second step is to construct two and three-body
representations by taking tensor products of one-body irreducible
representations.  The third step is to use Poincar\'e Clebsch-Gordan
coefficients in the chosen basis to reduce the tensor product of
irreducible representations to a direct integral of irreducible
representations.  The final step is to add an interaction to the mass
Casimir operator of the non-interacting irreducible representation
that commutes with and is independent of the linear momentum and
z-component of the canonical spin.  Simultaneous eigenstates of the
mass, linear momentum, spin, and z-component of the canonical spin are
constructed by diagonalizing the interacting mass operator in the
free-particle irreducible basis.  These eigenstates are complete and
transform irreducibly, and thus define the dynamical representation of
the Poincar\'e group.

The invariance principle implies the equality of two-body wave
operators constructed with the non-relativistic Hamiltonian $H= P^2
/4M + k^2/m + v_{NN}$ and the relativistic mass square operator $M^2 =
4 (k^2 + m^2) + 4mv_{NN}$ provided the relative momentum, $k$,
obtained by transforming the single-particle momentum to the rest
frame with a Galilean boost is identified with the relative momentum
obtained by transforming the single-particle momentum to the rest
frame with an inverse Lorentz boost.  Because of this identification,
the relativistic and non-relativistic two-body scattering matrices are
identical functions of $k$

Differences occur in how the two-body interactions appear in the 
three-body mass (rest energy) operator.  In the relativistic case the
non-linear relation between mass and energy, which must be respected
for S-matrix cluster properties, implies that the two-body 
interaction in the three-body problem has the form
\[
V_{12} = \sqrt{4k^2 + 4m^2 +
4mv_{NN} + q^2} - \sqrt{4k^2 + 4 m^2  + q^2}
\]
where $q$ is the 
momentum of the spectator boosted to the rest frame of the 
non-interacting three-body system.  The two-body interaction in the 
square root appears in the  
kernel of the Faddeev equation for the mass operator.

The relation between $M^2$ and the non-relativistic Hamiltonian leads
to the following identification for the non-relativistic two-body and
relativistic 2+1-body right-half-shell transition matrix elements
\[
\langle q', k' \vert T_{12}(z_r ) \vert q ,k \rangle =
{ 4m \langle  k' \vert t_{12 nr}(z_{nr}) \vert k \rangle \over 
\sqrt{4k^2 + 4 m^2  + q^2} + \sqrt{4k^{\prime 2} + 4m^2 + q^2}
}
\]
where $z_r$ and $z_{nr}$ are the right-half-shell invariant mass
squared and energy respectively.  The off-shell $T_{12}(z)$ is
needed in the Faddeev kernel.  We calculate this using the first
resolvent equation which gives an integral equation for the off-shell
$T_{12}(z)$ using the half-shell $T_{12}(z_r)$ as input \cite{2}.

This method is used to solve the Faddeev equations for the Poincar\'e
invariant mass operator.  In addition, because we are interested in
energies near the GeV scale, we have chosen to solve the equations by
direct integration, rather than dealing with the large number of
partial waves required for convergence \cite{3}\cite{4}.

Our test calculations, using a Malfliet-Tjon type of two-body 
interaction, are converged for laboratory energies up to 
2 GeV.    

While the model interaction does not have the spin complexities of a
realistic interaction, the success of these calculations suggests that
the method outlined above is suitable for modeling reactions in the
few GeV energy range.   Note that similar calculations using partial 
wave methods with realistic interactions have been solved at lower
energies \cite{5}.

\section{Results}

Comparison of the relativistic and corresponding non-relativistic
calculations lead to observations that should also be relevant for
realistic interactions.  Note that this comparison does not involve a
non-relativistic limit, instead relativistic and non-relativistic
three-body calculations with interactions that are fit to the same
two-body data are compared.  All of the differences are due to the
different ways in which the two-body dynamics appears in the three-body
problem.

One important observation is that the effect of relativistic
kinematics, which can be large and grows with energy, is largely
canceled by dynamical effects.

For elastic scattering the largest difference between the relativistic
and non-relativistic calculations is at large angles, above 
60 degrees at 1 GeV; the difference increases as a function of beam energy.

The convergence of the multiple scattering series is not uniform.  For
elastic scattering the first order term is not adequate even at 2 GeV 
for large angles.

Similar behavior is observed for breakup calculations away from the
quasielastic peaks.  In addition, both inclusive and exclusive breakup
calculations show a shift in the position of the quasielastic peaks
relative to the peaks in the non-relativistic theory.


\begin{thebibliography}{3}

\bibitem{1}
E.~P.~Wigner,
Ann. Math.  {\bf 40}, 149 (1939).

\bibitem{2}
B.~D.~Keister and W.~N.~Polyzou,
Phys. Rev.  C {\bf 73}, 014005 (2006).

\bibitem{3}
H.~Liu, Ch.~Elster, and W.~Gl\"ockle, Phys. Rev. C{\bf 72}, 054003 ( 2005).

\bibitem{4}
T.~Lin, C.~Elster, W.~N.~Polyzou and W.~Gl\"ockle,
Phys. Rev. C {\bf 76}, 014010 (2007).

\bibitem{5} H.~Wita{\l}a, J.~Golak, W.~Gl\"ockle, and H.~Kamada, Phys. Rev. 
C{\bf 71}, 054001 (2005).

\end{thebibliography}
\end{document}